 \documentclass[10pt,journal]{IEEEtran}

\usepackage[T1]{fontenc}      
\usepackage[utf8]{inputenc}   
\usepackage[british,UKenglish,USenglish,american]{babel}
\usepackage{csquotes}
\usepackage[backend=bibtex,style=ieee,maxcitenames=10,mincitenames=1,maxnames=10,minnames=1]{biblatex}
\usepackage[pdftex]{graphicx}
\usepackage{etoolbox}
\usepackage[ruled,vlined,linesnumbered]{algorithm2e}
\usepackage{url}
\usepackage{xcolor}
\usepackage{easyReview}
\usepackage{float}
\usepackage{multirow}
\usepackage[cmex10]{amsmath}
\usepackage{array}

\newcolumntype{L}[1]{>{\raggedright\let\newline\\\arraybackslash\hspace{0pt}}m{#1}}
\newcolumntype{R}[1]{>{\raggedleft\let\newline\\\arraybackslash\hspace{0pt}}m{#1}}
\newcolumntype{C}[1]{>{\centering\let\newline\\\arraybackslash\hspace{0pt}}m{#1}}

\usepackage{lipsum}
\usepackage{lscape}
\usepackage{epstopdf}
\usepackage{dblfloatfix}
\usepackage{color}
\usepackage{comment}
\usepackage{subcaption}
\usepackage{array}
\usepackage{setspace}

\AppendGraphicsExtensions{.tiff}

\begin{document}

\newcommand{\missref}{\alert{[REF]}}
\makeatletter
\def\missref{\@ifnextchar[{\@with}{\@without}}
\def\@with[#1]{\alert{[REF]}\foreach \n in {2,...,#1}{\alert{,~[REF]}}}
\def\@without{\alert{[REF]}}
\makeatother

\graphicspath{{./figures/}}


\title{A Two-Level Approximate Logic Synthesis Combining Cube Insertion and Removal}

\newbool{journalBasedAuthors}
\ifCLASSOPTIONpeerreview \setbool{journalBasedAuthors}{true}
\else \ifCLASSOPTIONjournal \setbool{journalBasedAuthors}{true}
\else \setbool{journalBasedAuthors}{false}
\fi
\fi

\ifbool{journalBasedAuthors}
{

    \author{Gabriel~Ammes, 
            Walter~Lau Neto, 
            Paulo~Butzen,
            Pierre-Emmanuel~Gaillardon, 
            Renato~P.~Ribas
    \thanks{G. Ammes and R. P. Ribas are with the Institute of Informatics, and P. Butzen is with School of Engineering, Federal University of Rio Grande do Sul (UFRGS), Porto Alegre, Brazil (e-mails: gabriel.ammes@inf.ufrgs.br, renato.ribas@ufrgs.br, paulo.butzen@ufrgs.br).}
    \thanks{W. Lau Neto and P-E. Gaillardon are with the Department of Electrical and Computer Engineering, University of Utah, Salt Lake City, US (e-mails: walter.launeto@utah.edu, pierre-emmanuel.gaillardon@utah.edu).}
    \thanks{This study was partially financed by the Brazilian funding agencies CAPES and CNPQ, and by DARPA, under the grant \# FA8650-18-2-7849.}
    \thanks{This work has been submitted to the IEEE for possible publication. Copyright may be transferred without notice, after which this version may no longer be accessible.}
    }
}

\markboth{}{Ammes \MakeLowercase{\textit{et al.}}: A Two-Level Approximate Logic Synthesis Combining Cube Insertion and Removal}


\maketitle

\begin{abstract}

Approximate computing is an attractive paradigm for reducing the design complexity of error-resilient systems, therefore improving performance and saving power consumption. In this work, we propose a new two-level approximate logic synthesis method based on cube insertion and removal procedures. Experimental results have shown significant literal count and runtime reduction compared to the state-of-the-art approach. The method scalability is illustrated for a high error threshold over large benchmark circuits. The obtained solutions have presented a literal number reduction up to 38\%, 56\% and 93\% with respect to an error rate of 1\%, 3\% and 5\%, respectively.

\end{abstract}


\ifCLASSOPTIONpeerreview
\vspace{.5cm}
Dear Editors.\vspace{.25cm}

The cover Letter comes here.
\fi

\IEEEpeerreviewmaketitle

\section{Introduction}

In the last decades, the complexity of electronic systems has grown very fast, impacting the circuit power dissipation, performance and area (PPA). Meanwhile, widely-used applications, such as signal processing, machine learning and data mining, exhibit error resilience properties. In this context, approximate computing has received special attention as a new design paradigm in recent years \cite{paper:surveyXu}\cite{paper:surveyMittal}. Such a paradigm consists in modifying the functionally behavior of digital circuits to reduce PPA. When an approximate circuit is applied to an error-resilient application, the error introduced tends to be not so critical to the final operation, and improvements on PPA are expected. Particular effort has been made over adders and multipliers through handcrafted designing and systematic synthesis of such a regular arithmetic structure \cite{paper:DATE2017}\cite{paper:Integration2020}. 

This work exploits approximate logic synthesis (ALS), which automatically synthesizes approximate circuits for specified Boolean functions \cite{paper:surveyALS}. ALS approaches for two-level (2L) and multilevel combinational circuits as well as for sequential logic design have been presented in the literature \cite{paper:DATE10,paper:ICCAD13, paper:ASICON15, paper:TCAD20, paper:ALSRAC, paper:seqALS}. In particular, two-level circuit synthesis consists in modifying a sum-of-products (SOP) expression of a given Boolean function aiming to minimize the literal count \cite{paper:DATE10,paper:ICCAD13, paper:ASICON15, paper:TCAD20}. It plays an essential role in the digital circuit synthesis over CPLD architectures \cite{paper:cpld}, as well as in multilevel logic synthesis \cite{paper:2l-ml}.

\IEEEpubidadjcol

The main goal of 2L-ALS methods is identifying an approximate SOP expression with the fewest number of literals for a given original SOP and error threshold.
This work applies the error rate (ER) metric as error constraint. The ER metric represents the probability that a given input vector leads to an erroneous output signal.
In \cite{paper:DATE10}, the authors present two techniques to approximate SOP: the insertion of cubes into the Boolean expression, by flipping the output from 0 to 1, and the removal of cubes from the expression, by switching the output from 1 to 0. They carried out experiments comparing both strategies and concluded that the cube insertion into the SOP leads to better results than the cube removal procedure. Hence, based on this assumption, related works have preferred the cube insertion to approximate SOP expressions.

In this work, we present a new 2L-ALS method that exploits cube insertion and removal by considering simultaneously both strategies without significant penalty in computation. At first, the proposed approach applies a cube insertion procedure, similar to those presented in \cite{paper:DATE10} and \cite{paper:TCAD20}, to provide partial SOP solutions with fewer errors than the specified ER threshold. In the following, a cube removal procedure is applied over the obtained SOPs, taking into account the remaining error slack.

The major contribution of our approach is to exploit both cube insertion and removal procedures together in a unified 2L-ALS method. The experiments carried out over benchmark circuits, with a threshold of 16 errors, provided results with 8\% fewer literals on average than the state-of-art method, without penalty in execution time \cite{paper:TCAD20}. Moreover, with a given ER percentage threshold, it reduced up to 38\% and 96\% in the literal count to an ER threshold of 1\% and 5\%, respectively.
It is worth emphasizing that, for the biggest benchmark circuit applied, the ER threshold of 5\% comprised up to 6,553 errors introduced. It illustrates the method scalability, knowing that, in \cite{paper:TCAD20}, the error insertion is at most 16 for the same benchmark.

The rest of this paper is organized as follows. Section \ref{section:preliminaries} presents some fundamentals, including the adopted terminology. Section \ref{section:related} discusses Su's work, taken here as the reference 2L-ALS method \cite{paper:TCAD20}. The proposed 2L-ALS method is described in Section \ref{section:method}, whereas experimental results are provided in Section \ref{section:results}. Finally, Section \ref{section:conclusion} concludes this paper.

\vspace{-1.5ex}
\section{Preliminaries}
\label{section:preliminaries}

In this section, the fundamentals on approximate logic synthesis and error metrics are briefly reviewed. The adopted terminology is also presented for a better understanding of the proposed 2L-ALS method.

A \textit{variable} corresponds to the symbol used to represent input and output signals. An occurrence of a variable in the Boolean expression is called a \textit{literal}. It can represent an input or an output literal, being that the input literal can be direct or complemented. A product of literals where any variable appears at most once is a \textit{cube}, and the sum of cubes results in a 
\textit{sum-of-products} (SOP). The particular case when a cube comprises a single literal for each input variable and only one output literal is called a \textit{minterm}. The size of a given cube is equal to the number of minterms it covers, whereas the expansion of a cube corresponds to the removal of one of the literals, turning it into a larger cube.

In approximate circuit design, several metrics have been adopted to quantify and restrict the error introduced \cite{paper:metrics}. Our approach applies the \textit{error rate} (ER) metric for restraining the error occurrence. The ER metric corresponds to the ratio between input combinations that leads to output errors and the total input vectors allowed. It is also referred to as the probability of error occurring for a given input.

In terms of approximate logic synthesis, an input combination that results in one or more outputs with incorrect value is defined as an \textit{erroneous input combination} (EIC). For instance, if two erroneous minterms present identical input literals but different output ones, only a single EIC is taken into account. Therefore, the \textit{number of errors} (NoE) in a given SOP is equal to its number of EICs. When using NoE as the error threshold, it can be an arbitrary value or equal to $2^n * er$, where $n$ is the number of inputs and $er$ corresponds to the ER threshold.

\vspace{-1.5ex}

\section{Related Work}
\label{section:related}

In \cite{paper:TCAD20}, Su \textit{et al.} present a heuristic search method to solve the 2L-ALS problem taking into account ER constraint. This work can be considered as the state-of-the-art method in the subject, being presented in the following.

The main goal of the Su's approach is to identify the set of input combinations for 0-to-1 output complement (SICC) that maximize the literal count reduction on an approximate SOP. It is similar to selecting the set of EICs that results in the most compact SOP. They propose an SICC-cube tree (SCT) data structure, which groups a set of EICs to a set of cubes that depends on these EICs to be inserted into the SOP. It comprises a two-level tree where the root contains the EICs and the leaves represent the cubes to be added into the SOP. The number of EICs in the root is equal to the number of errors inserted into the SOP.

Two conditions must be satisfied to ensure that SCT leaves lead to the optimization of the literal count. Firstly, at least one cube must be removed from the SOP when a new cube is inserted. Secondly, the literal count in the removed cubes must be greater than the literals present in inserted cube.

Their initial task enumerates all possible multiple-output cubes of a function through the Hasse diagram structure. These cubes are used to build a set of SCTs. In the next task, the SCTs are combined because there are some with fewer errors than the maximum number allowed. After that, it is necessary to select the SCT that reduces the greatest number of literals.

A straightforward way to calculates the literal reduction in a given SCT is by using the Espresso tool \cite{book:espresso}, taking into account the EICs on the root as \textit{don't cares} to obtain an approximate SOP. The calculation of the literal reduction with Espresso presents a precise result, but the impact on the runtime is quite significant. Hence, a procedure that avoids the use of Espresso for estimating such a reduction is presented.

The procedure to predict the literal reduction on an SCT comprises main three steps. As the insertion of leaf cubes into the SOP does not guarantee a reduction in literal count, it first identifies the set of cubes that may be removed when the leaf cubes are inserted. Moreover, inserting all leaf cubes may increase the SOP literal count. Thus, it identifies the set of leaf cubes necessary to be inserted before removing the first set of cubes. Finally, it calculates the literal reduction between the sets of cubes removed and inserted.

In \cite{paper:TCAD20}, the authors present four speed-up techniques to extend the application of their approach to large circuits.
\begin{enumerate}
\item As the basic algorithm time complexity grows exponentially with the NoE, the errors allowed for each execution is limited to two, therefore generating partial approximate SOPs. All partial SOPs are approximated again until the accumulated NoE reaches the threshold allowed.
\item With the first speed-up technique, an exponential quantity of partial approximate SOPs is created, so impacting the final runtime. In order to reduce the number of partial SOPs, only the two expressions with the fewest number of literals are approximate again for a given partial NoE. 
\item To reduce the number of combined SCTs, only a subset of all generated SCTs are taken into account. At first, it estimates the literal count of all SCTs without combining them. In the next, for combining two SCTs, the first one must be within the 25\% of the SCTs with the fewest number of literals whereas the second must be within the 80\%.
\item The treatment of all cubes present on the Hasse diagram implies a high computational cost. In order to reduce such a cost, they only take the cubes on the diagram that are the parents of the cubes on the SOP, since it is improbable that any other than a parent of an SOP cube inserts less than two errors.
\end{enumerate}

The values used in the first three speed-up techniques were obtained through an empirical analysis. Nevertheless, there is still an important runtime bottleneck for large NoE.

\section{Proposed Two-Level ALS Method}
\label{section:method}

Existing 2L-ALS methods based on ER constraint have in common the adoption of cube insertion as the unique strategy adopted. On the other hand, the proposed approach exploits simultaneously both cube insertion and removal procedures, so aiming to reduce the literal count of a given optimized Boolean formulation $\mathcal{F}$ and ER threshold $er$. First of all, a general description of the 2L-ALS method is presented, and the algorithms to approximate the SOP are described in the following.

The related works do not present good execution scalability. For instance, the approaches presented in \cite{paper:DATE10} and in \cite{paper:TCAD20} limit the NoE to 8 and 16, respectively, due to the resulting runtime. The time complexity analysis of the proposed method is discussed at the end of this section.

\vspace{-1.5ex}
\subsection{General Description} \label{sc:overview}

At first, the method applies a cube insertion procedure and generate multiple partial approximate SOPs with at most a given NoE constraint. For these partial SOPs, a cube removal procedure is then applied to approximate them assuming the remaining NoE. The cube insertion procedure is based on the approach presented in \cite{paper:TCAD20}, and described in Section \ref{section:related}.

The overall flow is illustrated in Algorithm \ref{alg:geral}. The SOP expression is stored using two maps. The first map groups each SOP cube with the minterms only covered by this cube. The second one groups each minterm covered by the SOP with the cubes that cover it. The original SOP $\mathcal{F}$ is approximated taking into account an NoE threshold equal to 2. Two partial solutions comprising one and two errors are generated by applying the cube insertion algorithm. This procedure is based on the first speed-up technique presented in \cite{paper:TCAD20}. For each partial SOP solution, the cube removal procedure is applied considering the remaining slack NoE, which corresponds to the difference between the already inserted errors and the initial error constraint. Such a strategy helps in escaping from local minimum, thus leading to better solutions.

\vspace{-1.5ex}
\begin{algorithm}[ht]
\setstretch{0.8}
\SetAlgoLined
\KwIn{SOP expression $\mathcal{F}$ and ER threshold \textit{er}}
\KwOut{approximated SOP expression $\mathcal{F}$'}
 \textit{e} $\gets er * 2^n$ \;
 Set sols with \textit{e}+1 sets of solutions\; \nllabel{line:solutions}
 \textit{sols}\textsubscript{0} $\gets \oslash$ (empty solution)\;
 \For{$i\gets 0$ to \textit{e}}{
    \textit{topS} $\gets $ the two best solutions in \textit{sols}\textsubscript{i}\; \nllabel{line:top}
    \For{each solution \textit{s} in \textit{topS}}{
        modifySOP($\mathcal{F}$, \textit{s})\; \nllabel{line:ModifySop}
        (\textit{s1}, \textit{s2}) $\gets$ cubeInsertion($\mathcal{F}$, 2, \textit{s})\; \nllabel{line:ALSInsert}
        \textit{sols}\textsubscript{i+1} $\gets$ \textit{sols}\textsubscript{i+1} $\cup$ \textit{s1}\; \nllabel{line:PartialSolution1}
        \textit{sols}\textsubscript{i+2} $\gets$ \textit{sols}\textsubscript{i+2} $\cup$ \textit{s2}\; \nllabel{line:PartialSolution2}
        \textit{s3} $\gets$ cubeRemoval($\mathcal{F}$, \textit{e-i}, \textit{s})\; \nllabel{line:ALSRemove}
        \textit{sMax} $\gets$ max(\textit{s1}, \textit{s2}, \textit{s3})\;
        \lIf{sMax $>$ best}{\textit{best} $\gets$ \textit{sMax}}
        restoreSOP($\mathcal{F}$, \textit{s})\; \nllabel{line:RestoreSop}
    }
 }
 return espresso(modifySOP($\mathcal{F}$,\textit{best}))\;
 \caption{Proposed 2L-ALS Method}
 \label{alg:geral}
\end{algorithm}
\vspace{-1.5ex}

To prevent a critical increase in runtime and space complexity, only the initial SOP expressions are stored, and the partial solutions modify the original function $\mathcal{F}$. Therefore, a partial solution comprises the set of cubes to be inserted and removed, along with the literal reduction and EICs added. In line \ref{line:solutions}, the set $sols$ comprises all the partial solutions obtained, whereas the set of $sols_i$ comprises the partial solutions with $i$ errors inserted. As the set $sols_0$ contains zero errors, it is initialized with an empty solution. For each set $sols_i$, the two best solutions are selected in line \ref{line:top}, similar to the second speed-up technique presented in \cite{paper:TCAD20}.

To modify the SOP $\mathcal{F}$, taking into account a solution $s$, it is applied the modifySOP function, corresponding to line \ref{line:ModifySop}, in Algorithm \ref{alg:geral}, which inserts and removes cubes in $\mathcal{F}$.
Meanwhile, the restoreSOP function in line \ref{line:RestoreSop}, undoes these modifications. 
The cube insertion procedure, in line \ref{line:ALSInsert}, returns two solutions (\emph{s1} and \emph{s2}) that are stored in $sols_{i+1}$ and $sols_{i+2}$, in lines \ref{line:PartialSolution1} and \ref{line:PartialSolution2}. 
In the cube removal procedure, in line \ref{line:ALSRemove}, for a given NoE $e$, it is allowed to insert $e-i$ errors, returning the solution \emph{s3}.
The best solution is then updated with the solution that comprises the greatest literal count reduction.
At the end, the best solution is then applied to $\mathcal{F}$ and optimized by the Espresso tool \cite{book:espresso}.
\vspace{-1.5ex}

\subsection{Cube Insertion Procedure}

The cube insertion procedure is based on the heuristic search method presented in \cite{paper:TCAD20}. It uses the SICC-cube tree (SCT) as the primary data structure to perform the approximation. The main idea is to generate SCTs from cubes that do not exceed the threshold of EIC number, and then select the SCT with the most significant literal reduction.
It is worth mentioning that, since our NoE is equal to 2, the SCT root has two EICs at most.

\vspace{-1.5ex}
\begin{algorithm}[ht]
\setstretch{0.8}
\SetAlgoLined
\KwIn{a simplified SOP expression $\mathcal{F}$, an NoE threshold \textit{e}, and the actual solution \textit{s}}
\KwOut{two solutions with error 1 and 2}
 \textit{trees} $\gets$ generateSCT($\mathcal{F}$, \textit{e}, \textit{s}.EIC)\; \nllabel{line:GenerateSCT}
 augment(\textit{trees})\; \nllabel{line:augment}
 (\textit{s1}, \textit{s2}) $\gets$ combineAndEstimate($\mathcal{F}$, \textit{trees})\; \nllabel{line:combineAndEstimate}
 \textit{s1} $\gets$ updateSolution(\textit{s1}, \textit{s})\;\nllabel{line:update1}
 \textit{s2} $\gets$ updateSolution(\textit{s2}, \textit{s})\;\nllabel{line:update2}
 return (\textit{s1}, \textit{s2})\;
 \caption{cubeInsertion Procedure}
 \label{alg:insert}
\end{algorithm}
\vspace{-1.5ex}

Algorithm \ref{alg:insert} presents the cube insertion flow. 
In line \ref{line:GenerateSCT}, it generates all SCTs.
The expanded cubes from $\mathcal{F}$ are considered as possible leaves to generate the SCTs. Using the expanded cube simplifies the third speed-up technique, mentioned in Section \ref{section:related}, as the Su's approach can insert any parent cube of an SOP cube. 
When an expanded cube is used as a possible SCT leaf, it is guaranteed the removal of the originating cube results in the literal count reduction. That way, it only has to verify the number of EICs needed to insert an expanded cube.
The EICs of a cube comprise all minterms covered by it that are not covered by $\mathcal{F}$, and its input combination that was not previously added as an EIC.

For SCTs \emph{sct1} and \emph{sct2} comprising one and two EICs in the root, if the \emph{sct2} root contains the \emph{sct1} root, the leaves of the \emph{sct1} are inserted into the \emph{sct2} leaves.
This updated \emph{sct2} is called an augmented SCT, as seen in line \ref{line:augment}.

In line \ref{line:combineAndEstimate}, the combination of SCTs and the estimation of literal reduction are performed.
At first, the literal reduction of all generated SCTs is estimated. 
For all SCTs with one EIC in the root, their roots and leaves are combined two by two through the fourth speed-up technique presented in Section \ref{section:related}.
The two solutions that reduce more literals with NoE equal to 1 and 2 are returned and stored in \emph{s1} and \emph{s2}.

The solutions \emph{s1} and \emph{s2} are updated in lines \ref{line:update1} and \ref{line:update2}.  
This update comprises the following steps: adding the cubes inserted and removed within solution \emph{s} into solutions \emph{s1} and \emph{s2}; estimating the new literal reduction; and updating the EICs. 
At the end, this procedure returns the solutions \emph{s1} and \emph{s2}.
\vspace{-1.5ex}

\subsection{Cube Removal Procedure}

Removing a cube implies that the cube literals are removed from the SOP. Therefore, the cube removal procedure is a greedy algorithm that selects the cube with the largest ratio between the numbers of literals and EICs.
Algorithm \ref{alg:remove} shows the flow of this procedure.

In the loop presented in line \ref{line:inicioFor}, the cube is chosen for removal. 
In this case, the EICs are obtained in line \ref{line:getEIC} and the gain in line \ref{line:gain}.
While removing a cube, its EICs are given by the minterms covered only by this cube in $\mathcal{F}$ whose input combination was not previously added as an EIC. 
Those minterms are in the first map of the SOP data structure.
In cases where the gain is greater than the actual bestGain, the gain, the cube and its EICs are stored.
If a cube has no EICs, its gain is maximized.
In line \ref{line:removeFromSop}, the best cube is removed from $\mathcal{F}$.
When the removal task is done, the minterms only covered by each cube in $\mathcal{F}$ are updated, impacting the subsequent iterations.
Then, the set of the removed cubes and the NoE are updated.
The updateEIC procedure, in line \ref{line:eic}, updates the newEICs set by adding on it the EICs of the bestEIC, and verifying whether there are EICs that were corrected by the cube removal procedure.

\vspace{-1.5ex}
\begin{algorithm}[ht]
\setstretch{0.8}
\SetAlgoLined
\KwIn{a simplified SOP expression $\mathcal{F}$, an NoE threshold \textit{e}, and the actual solution \textit{s}}
\KwOut{a solution with at most \textit{e} errors}
 \textit{error} $\gets$ \textit{e}, \textit{newEIC} $\gets$ \textit{s}.EIC\;
 \While{error $> 0$}{
    \For{each Cube in $\mathcal{F}$}{\nllabel{line:inicioFor}
        \textit{cubeEIC} $\gets$ getCubeEIC(\textit{cube}, $\mathcal{F}$, \textit{newEIC})\; \nllabel{line:getEIC}
        \textit{gain} $\gets$ litCount(\textit{cube}) / max(0.01, \#\textit{cubeEIC})\; \nllabel{line:gain}
        \If{gain $>$ bestGain and error $\geq$ \#\textit{cubeEIC}}{
            \textit{bestGain} $\gets$ \textit{gain}\;
            \textit{bestCube} $\gets$ \textit{cube}\;
            \textit{bestEIC} $\gets$ \textit{cubeEIC}\;
        }
    }\nllabel{line:fimFor}
    removeCubeFromSOP(\textit{bestCube}, $\mathcal{F}$)\; \nllabel{line:removeFromSop}
    \textit{removedCubes} $\gets$ \textit{removedCubes} $\cup$ \textit{bestCube}\;
    \textit{error} $\gets$ \textit{error} - \#\textit{bestEIC}\;
    \textit{newEIC} $\gets$ updateEICs(\textit{newEIC}, \textit{bestEIC})\;\nllabel{line:eic}
 }
 insertCubes($\mathcal{F}$, \textit{removedCubes})\; \nllabel{line:insertInSop}
 \textit{s3} $\gets$ updateSolution(\textit{removedCubes}, \textit{newEic}, \textit{s})\;
 return \textit{s3}\;
 \caption{cubeRemoval Procedure}
 \label{alg:remove}
\end{algorithm}

When it reaches the allowed NoE, the main loop ends.
At the end, the cubes are re-inserted in $\mathcal{F}$, and the solution s3, comprising the removed cubes, the inserted EICs and an updated literal reduction count, is returned.

\vspace{-2.5ex}

\subsection{Time Complexity Analysis}

The cubeInsertion and cubeRemoval procedures present the most relevant impact on the time complexity of the proposed method. Each of them is executed \emph{e} times. For the sake of simplicity, we are omitting the Espresso complexity.

The combineAndEstimate task in the cubeInsertion procedure is the most time consuming one.
The most expensive phase of this procedure is to combine two by two the SCTs with one EIC on the root and estimate their literal count reduction. 
To generate the SCTs, the cubes on the SOP are expanded.
As the expansion generates a new cube for every literal in a cube, the number of expanded cubes is equal to the number of literals in the SOP, represented by \emph{L}.
The worst-case number of SCTs with one EIC is reached when each expanded cube generates one of them. 
Thus, the number of SCTs that have their literals estimated is up to $O$($L^2$). The literal estimation depends on obtaining the covered minterms of each leaf cubes. As the number of covered minterms by a cube is at most $O$($m*2^n$), where $n$ and $m$ are the number of inputs and outputs of the function, respectively, the worst-case time complexity of the cubeInsertion procedure is $O$($L^2*m*2^n$).

The cubeRemoval procedure, in turn, estimates the gain of removing each cube, represented by \emph{C}, and removes the one with more gain until the limit error is reached.
The gain depends on the number of EICs and cube literals. As obtaining the EICs relies on hash structures, its time complexity can be taken as constant. To obtain literal count, the cube are iterated $O$($n+m$) times. Therefore, the worst-case time complexity of the cubeRemoval procedure is $O$($e*C*$ ($n+m$)). 

The complete worst-case time complexity of the proposed method is $O$($e*$($L^2*m*2^n$ + $e*C$ ($n+m$))).

\vspace{-1ex}

\section{Experimental Results}
\label{section:results}

The proposed algorithms have been implemented in the C++ programming language.
Our experiments have been carried out over the IWLS'93 benchmark suite \cite{paper:IWLS93}, in a computer with a quad-core i5-2400 CPU @ 3.10GHz and 8GB of RAM.

\vspace{-1.5ex}

\subsection{Comparison to other approaches}

The Su's approach, presented in \cite{paper:TCAD20}, is the state-of-the-art 2L-ALS published method, so it has been taken into account herein as our golden reference. 
The experiments consider the same circuits as in \cite{paper:TCAD20}, with the NoE threshold equals to 16 in order to allow a fair comparison. Therefore, the designs have more than 6 and fewer than 20 inputs, and the sum of inputs and outputs has less than 34. As Su's approach source code is not publicly available, we are comparing our results with the ones presented in \cite{paper:TCAD20}, whose experiments were carried out with a quad-core i5-6500 CPU @ 3.20GHz and 32GB of RAM.

Table \ref{tab:16} shows the comparison results between Su's method and our proposed approach. Column 1 presents the name of the circuits, as well as the number of inputs (\textit{i}) and the outputs (\textit{o}). 
Column 2 shows the number of literals of the original SOPs, whereas
column 3 and column 4 present the number of literals of the approximate circuits presented in \cite{paper:TCAD20} and obtained by our method, respectively. 
Column 5 and column 6 show the literal reduction rate between the literal count of the original SOP and our approximate solution, and between literal count from the approximate SOP generated by our approach and the one presented in \cite{paper:TCAD20}, respectively.
Column 7 and column 8 present the runtime for both methods. 

\begingroup
\setlength{\tabcolsep}{2pt} 
\renewcommand{\arraystretch}{1} 

\begin{table}[tpb]
\scriptsize
\caption{Comparison to Su's approach \cite{paper:TCAD20} in IWLS93 benchmark suite considering NoE threshold of 16.}
\label{tab:16}
\centering
\begin{tabular}{ |l|c|c|c|c|c|c|c|c| } 
\hline
\multirow{2}{*}{Circuit} & \multirow{2}{*}{ER} &  \multicolumn{5}{c|}{Literals} & \multicolumn{2}{c|}{Time(s)} \\ \cline{3-9}
& & Orig. & Su's & Ours & Red. & Ours/Su's & Su's & Ours \\ \hline 
con1 i:7;o:2 & 12.50\% & 32 & 32 & 24 & 0.75 & \textbf{0.75} & 0.38 & 0.02 \\ \hline
rd73 i:7;o:3 & 12.50\% & 903 & 578 & 556 & 0.61 & \textbf{0.96} & 1.48 & 0.92\\ \hline
inc i:7;o:9 & 12.50\% & 198 & 156 & 125 & 0.63 & \textbf{0.80} & 0.49 & 0.13\\ \hline
5xp1 i:7;o:10 & 12.50\% & 347 & 235 & 202 & 0.58 & \textbf{0.85} & 0.72 & 0.47\\ \hline
sqrt8 i:8;o:4 & 6.25\% & 188 & 98 & 83 & 0.44 & \textbf{0.84} & 0.58 & 0.16\\ \hline
rd84 i:8;o:4 & 6.25\% & 2070 & 1578 & 1511 & 0.72 & \textbf{0.95} & 6.52 & 3.03\\ \hline
misex1 i:8;o:7 & 6.25\% & 96 & 96 & 77 & 0.80 & \textbf{0.80} & 0.50 & 0.01\\ \hline
clip i:9;o:5 & 3.12\% & 793 & 588 & 584 & 0.73 & \textbf{0.99} & 1.99 & 0.95\\ \hline
apex4 i:9;o:19 & 3.12\% & 5419 & 5040 & 5024 & 0.92 & \textbf{0.99} & 109 & 22.08\\ \hline
sao2 i:10;o:4 & 1.56\% & 496 & 231 & 165 & 0.33 & \textbf{0.71} & 2.48 & 1.04\\ \hline
ex1010 i:10;o:10 & 1.56\% & 2718 & 2693 & 2636 & 0.96 & \textbf{0.97} & 14.30 & 1.46\\ \hline
alu4 i:14;o:8 & 0.09\% & 5087 & 4904 & 4847 & 0.95 & \textbf{0.98} & 298 & 9.73\\ \hline
misex3 i:14;o:14 & 0.09\% & 7784 & 7446 & 7242 & 0.93 & \textbf{0.97} & 693 & 8.08\\ \hline
table3 i:14;o:14 & 0.09\% & 2644 & 2459 & 2347 & 0.88 & \textbf{0.95} & 513 & 3.77\\ \hline
misex3c i:14;o:14 & 0.09\% & 1561 & 1239 & 1115 & 0.71 & \textbf{0.89} & 252 & 13.35\\ \hline
b12 i:15;o:9 & 0.04\% & 207 & 207 & 207 & 1.00 & 1.00 & 249 & 1.14\\ \hline
t481 i:16;o:1 & 0.02\% & 5233 & 5105 & 4975 & 0.95 & \textbf{0.97} & 1570 & 2.25\\ \hline
table5 i:17;o:15 & 0.01\% & 2501 & 2410 & 2270 & 0.90 & \textbf{0.94} & 7868 & 16.08\\ \hline
Average & - & 2126 & 1949 & 1888 & 0.76 & \textbf{0.90} & 643 & 4.69\\ \hline

\end{tabular}
\vspace{-4ex}
\end{table}

\endgroup

Our method has showed better results for all benchmark circuits treated, except the \emph{b12} one that both approaches were not able to optimize.
The circuits \emph{con1}, \emph{misex1} and \emph{b12} could not be approximated by Su's method as it does not have SCT with size equals to one or two.
On the other hand, \emph{con1} and \emph{misex1} have been approximated by our method due to the cube removal phase.

Moreover, the proposed method presented a better efficiency in general, with average runtime of around 4.69s in comparison to 643s presented in \cite{paper:TCAD20}.
Such a difference is observed for circuits with more than ten inputs, where our method has a more scalable temporal behavior.

\vspace{-1.5ex}

\subsection{Results with error rate}

Fixing an NoE threshold may be a problem because the ER depends on the number of input combinations related to the target circuit.
For instance, the 16 NoE applied before corresponds to an ER of 12,5\% for a circuit with 7 inputs but 0.01\% for a circuit with 17 inputs.
As our method presents a good runtime efficiency, it has been possible to apply a higher NoE and consequently allowed us to use a percentage error rate.
Table \ref{tab:er} shows the approximate SOP solutions considering an ER of 1\%, 3\% and 5\% over benchmark circuits with more than 10 inputs.
Column 1 gives the circuits and their input and output numbers. 
Column 2 and column 3 provide the ER in percentage and the corresponding NoE, respectively.
Column 4 presents the literal count of the approximate SOPs obtained from our method.
Column 5 shows the literal count reduction in percentage, whereas column 6 provides the runtime.

Our method reaches an average literal reduction of 38\% with ER of 1\%, 56\% with ER of 3\%, and 64\% with an ER of 5\%.
For \emph{sao2}, \emph{table3}, \emph{t481} and \emph{table5} circuits, we obtained a literal count reduction close to 90\% with an ER of 5\% and up to 93.9\% with the same ER for the \emph{table5} circuit.
Moreover, even though the \emph{b12} could not be approximated with a 16 NoE, we have shown that while considering an ER percentage as a constraint, it can be approximated with a literal count reduction up to 26.1\%. Even with a higher NoE in circuits with many variables, the runtime remains under 5 minutes.

\begingroup
\setlength{\tabcolsep}{5pt} 
\renewcommand{\arraystretch}{1} 

\begin{table}[ht]
\caption{Result of the proposed method considering ER threshold in IWLS93 benchmark suit.}
\label{tab:er}
\centering
\scriptsize
\begin{tabular}{ |c|c|c|c|c|c| } 
\hline
\multirow{2}{*}{\centering Circuit} & \multirow{2}{*}{\centering ER} & \multirow{2}{*}{\centering NoE} & \multicolumn{2}{c|}{Literals} & \multirow{2}{*}{\centering Time (s)} \\ \cline{4-5}
 & & & Original & Approximate  & \\ \hline
sao2  & 1\%  & 10 & \multirow{3}{3.2em}{\centering 496} & 274 (0.55) & 0.56 \\ \cline{2-3}\cline{5-6}
i: 10  & 3\%  & 30 &  & 79 (0.15) & 2.16 \\ \cline{2-3}\cline{5-6}
o: 4  & 5\%  & 51 &  & 37 (0.07) & 3.23 \\ \hline
ex1010  & 1\%  & 10 & \multirow{3}{3.2em}{\centering 2718} & 2659 (0.97) & 0.90 \\ \cline{2-3}\cline{5-6}
i: 10  & 3\%  & 30 &  & 2588 (0.95) & 2.76 \\ \cline{2-3}\cline{5-6}
o: 10  & 5\%  & 51 &  & 2511 (0.92) & 4.80 \\ \hline
alu4  & 1\%  & 163 & \multirow{3}{3.2em}{\centering 5087} & 3730 (0.73) & 96.07 \\ \cline{2-3}\cline{5-6}
i: 14  & 3\%  & 491 &  & 2693 (0.52) & 188.18 \\ \cline{2-3}\cline{5-6}
o: 8  & 5\%  & 819 &  & 2139 (0.42) & 279.55 \\ \hline
b12  & 1\%  & 372 & \multirow{3}{3.2em}{\centering 207} & 193 (0.93) & 1.57 \\ \cline{2-3}\cline{5-6}
i: 15  & 3\%  & 983 &  & 170 (0.82) & 4.81 \\ \cline{2-3}\cline{5-6}
o: 9  & 5\%  & 1638 &  & 153 (0.73) & 10.08 \\ \hline
t481  & 1\%  & 655 & \multirow{3}{3.2em}{\centering 5233} & 1992 (0.38) & 3.14 \\ \cline{2-3}\cline{5-6}
i: 16  & 3\%  & 1966 &  & 942 (0.18) & 3.89 \\ \cline{2-3}\cline{5-6}
o: 1  & 5\%  & 3276 &  & 578 (0.11) & 5.92 \\ \hline
table5  & 1\%  & 1310 & \multirow{3}{3.2em}{\centering 2501} & 720 (0.28) & 100.98 \\ \cline{2-3}\cline{5-6}
i: 17  & 3\%  & 3932 &  & 280 (0.11) & 198.05 \\ \cline{2-3}\cline{5-6}
o: 15  & 5\%  & 6553 &  & 153 (0.06) & 245.94 \\ \hline

 \end{tabular}
\vspace{-4ex}
\end{table}

\endgroup

\section{Conclusions}
\label{section:conclusion}

This paper presented the first 2L-ALS method that exploits both insertion and removal of cubes as strategies to approximate a given SOP expression taking into account ER metric. Experimental results have shown that the proposed approach surpasses the state-of-the-art method in quality of results and scalability.
The related source code, the applied benchmark circuits used to generate the experimental results, and their approximate descriptions are publicly available on GitHub\footnote{\url{https://github.com/GabrielAmmes/2LALS-IR}}.


\AtNextBibliography{\small}

\printbibliography



\vfill

\newpage

\end{document}